\documentclass[11pt]{article}
\usepackage{latexsym,amssymb,amstext,amsmath,array}
\usepackage{amscd,bbm}
\usepackage{cite}
\usepackage{slashed}
\usepackage{mathrsfs}
\usepackage{hyperref}
\usepackage{xcolor}

\usepackage{xcolor}
\usepackage[english]{babel}
\usepackage[T1]{fontenc}
\usepackage[utf8]{inputenc}
\usepackage{authblk}
\usepackage{bm}
\usepackage{mathtools}
\usepackage{slashed}
\usepackage{enumerate}
\usepackage{graphicx,color}
\usepackage{cite}
\usepackage{float}
\usepackage{hyperref}
\hypersetup{
	colorlinks=true,
	linkcolor=blue,
	filecolor=magenta,      
	urlcolor=cyan,
}
\usepackage[capitalize]{cleveref}
\usepackage{verbatim}
\usepackage{lmodern}

\usepackage{setspace}

\allowdisplaybreaks
\graphicspath{{chobi/}}

\newcommand {\cN}{{\cal N}}

\def\G{\Gamma}

\newcommand{\be}{\begin{equation}}
\newcommand{\ee}{\end{equation}}
\newcommand{\bea}{\begin{eqnarray}}
\newcommand{\eea}{\end{eqnarray}}

\newcommand{\ba}{\begin{array}}
\newcommand{\ea}{\end{array}}

\def\double #1{#1{\hbox{\kern-2pt $#1$}}}

\newcommand{\bsubeq}{\begin{subequations}}
\newcommand{\esubeq}{\end{subequations}}

\begin{document}

\begin{titlepage}

\begin{center}

\vskip .3in \noindent

{\Large \bf{Equivalent Dual Theories for 3D \emph{$\cN$=2}  Supergravity }}

\bigskip

{Nabamita Banerjee}\footnote{nabamita@iiserb.ac.in (on lien from IISER Pune)}, {Saurish Khandelwal}\footnote{ks.saurish@gmail.com}, {Parita Shah}\footnote{paritars8@gmail.com}, \\ 

\bigskip
$^{a}$ \em Indian Institute of Science Education and Research Bhopal\\
Bhopal Bypass, Bhauri, Bhopal 462066, India \\

\vskip .5in
{\bf Abstract }
\vskip .2in
\end{center}

$\cN$=2 three dimensional Supergravity with internal $R-$symmetry generators can be understood as a two dimensional chiral Wess-Zumino-Witten model. In this paper, we present the reduced phase space description of the theory, which turns out to be flat limit of a generalised Liouville theory, up to zero modes. The reduced phase space description can also be explained as a gauged chiral Wess-Zumino-Witten model. We show that both these descriptions possess identical gauge and global (quantum $\cN$=2 superBMS$_3$) symmetries.

\vfill
\eject

\end{titlepage}

\newpage
\tableofcontents
\section{Introduction and Summary}
There is a connection between $D+1$ dimensional diffieomorphism invariant theories and $D$ dimensional field theories. Details of this duality strongly
depends on the precise form of boundary conditions on various fields. One of the simplest contexts where this has been studied is $2+1$ dimensional gravity theories. It is a well known fact that three dimensional gravity can be described by a two dimensional field theory. 3D gravity solutions with non-trivial topology correspond to stress-energy tensors of a dual two dimensional theory. This duality is best understood in the Chern-Simons formulation of 3D gravity \cite{WITTEN198846,Achucarro:1987vz}. 
The reduced dual theory in this case is in general a (chiral) Wess-Zumino-Witten (WZW) model\cite{witten1984},  defined on a closed spatial section and is
obtained by solving part of the constraints in the Chern-Simons theory\cite{Witten:1988hf,Moore:1989yh,Elitzur:1989nr}. Such reductions have been mostly performed for asymptotically AdS 3D gravity  \cite{Coussaert:1995zp,Floreanini:1987as,article,SONNENSCHEIN1988752,PhysRevLett.62.1817,Chu:1991pn,Caneschi:1996sr,Henneaux:1999ib,Valcarcel:2018kwd}, where the dual $2$D theory is a conformal field theory with infinite dimensional symmetry. In this paper, we are interested in the dual of asymptotically flat 3D (super)gravity.
In particular, ordinary asymptotically flat 3D gravity can be understood as a $ISO(2,1)$ Chern-Simons gauge theory with flat boundary condition at null infinity where the Chern-Simons level $k$ is identified with Newton's constant. Here the
spatial section is a plane and the choice of boundary conditions is crucial in determining the dual theory.  Reduction of  $ISO(2,1)$ Chern-Simons to WZW model was first studied in \cite{SALOMONSON1990769}.  An alternate route has been taken in \cite{Barnich:2013yka} where the dual WZW model has been constructed for flat ordinary 3D gravity\footnote{Higher spin and Supersymmetric  generalisations were performed in \cite{Afshar:2013vka,Gonzalez:2013oaa,Barnich:2014cwa,Lodato:2016alv,Banerjee:2017gzj,Fuentealba:2017fck,Basu:2017aqn,Banerjee:2018hbl}.}. In \cite{Barnich:2013yka}, other than $ISO(2,1)$ gauge algebra, the boundary conditions suitable for flat asymptotics (at null infinity) have been applied for the gauge field. As a result, the dual chiral WZW model, when is gauged, shows invariance under infinite dimensional quantum BMS$_3$ algebra which is the asymptotic symmetry of flat 3D gravity. The analysis was further extended for minimal ${\cal{N}}=1$ Supergravity theory in \cite{Barnich:2015sca}, higher spin gravity \cite{Gonzalez:2014tba}  and recently for ${\cal{N}}=2$ case, with(out) internal $R-$symmetry in \cite{Banerjee:2019lrv}.

In \cite{Barnich:2013yka,Barnich:2015sca}, it was further shown that for pure and ${\cal{N}}=1$ 3D supergravity theory, the asymptotic boundary conditions lead to a reduced phase space description as a flat limit of (super)Liouville theory at null infinity (up to zero modes).
In view of CS-WZW duality, we can understand this result as due to the fact that
the  asymptotic conditions are strong enough to enforce the
Hamiltonian reduction from SL(2,R)-WZW to Liouville theory \cite{Bershadsky1989,Alekseev:1988ce,ORaifeartaigh:1998pox}. Another way of looking at it would be to recall that the dual chiral WZW model shows further gauge invariance. It was described in \cite{PhysRevD.44.314,Witten:1991mk,witten1992,Ginsparg:1992af,FEHER19921} that particular sub-sectors of symmetry can be gauged without introducing any anomaly to the system. The gauged chiral WZW model then can be shown to be equivalent to the flat Liouville description. The gauging is identical to imposing first class constraints to WZW model, that arises due to the asymptotic boundary condition.

 The current paper should be considered as a follow up of our recent work \cite{Banerjee:2019lrv}, where we have constructed the dual chiral WZW model for ${\cal{N}}=2$ 3D supergravity with internal $R-$symmetry.
Here we present the reduced phase space description of the theory and study its properties. The reduced phase space turns out to be a  flat limit of a generalised superLiouville type theory and is identical to the dual  chiral WZW model constructed in \cite{Banerjee:2019lrv}, when appropriately gauged. Finally, We present the gauge invariance of the reduced system, which is same as the residual gauge invariance of the gauged chiral WZW model. 
 
 The paper it is organised as follows: in section \ref{sec2} we briefly present the ${\cal{N}}=2$ 3D supergravity with internal $R-$symmetry and its asymptotic boundary condition at null infinity that reproduces the infinite dimensional quantum BMS$_3$ symmetry. In section \ref{sec3}, we write down the equivalent chiral WZW model that describes the dynamics of the theory and present its symmetries with minimal required details. Then we present the gauged version of the theory. Section \ref{sec4} contains the main result of this paper, where we present the phase space description of the dual theory and show its equivalence with the gauged chiral WZW model. Section \ref{sec5} points an interesting outlook of this work. Our conventions and some computational details have been presented in the appendices.

\section{ $3$-dimensional  ${\cal{N}}=2$  Supergravity and its asymptotic symmetry}\label{sec2}
There are two different versions of ${\cal{N}}=2$ SuperPoincar\'{e} algebras known in the literature \cite{HOWE1996183}. One of it, commonly known as ${\cal{N}}=(1,1)$, contains two supercharges but no internal $R-$symmetry. The other one,known as ${\cal{N}}=(2,0)$ SuperPoincar\'{e} algebra, is more interesting as it allows  the two supercharges to transforms under an internal $R-$symmetry. The algebra can be presented as, 
\begin{eqnarray}\label{N=(2,0)}
[J_a, J_b] &= \epsilon_{abc} J^c~~~~~~
[J_a, P_b] = \epsilon_{abc} P^c\\ \nonumber
[J_a, Q^i_{\alpha}] &= \frac{1}{2} (\Gamma_a)^{\beta}_{\alpha} Q^i_{\beta}~~~~~~~~
[Q^i_{\alpha},T] = \epsilon^{ij} Q^j_{\alpha}\\ \nonumber
\{Q^i_{\alpha}, Q^j_{\beta} \} &= -\frac{1}{2} \delta^{ij} (C\Gamma^a)_{\alpha \beta} P_a + C_{\alpha\beta} \epsilon^{ij} Z. 
\end{eqnarray}	
Here $J_a, P_a  {(a= 0,1,2) }$ are the Poincare generators and $\mathcal{Q}_{\alpha}^{i}$,two distinct $({i=1,2})$ two component $({\alpha =+1,-1})$ spinors, are the two fermionic generators of the algebra. These fermionic generators transform under a spinor representation of an internal R-symmetry generator $T$. The above algebra has non-degenerate invariant bilinear only in presence of a central term $Z$  \cite{HOWE1996183}. Our conventions are presented in \ref{AppA}. In this paper, we shall work with a 3D supergravity theory invariant under the above symmetry. In the CS formulation, 3D (super)gravity theory can be represented as,
\begin{equation}\label{csaction}
I [A] = \frac{k}{4 \pi}\int_M \langle A, dA+ \frac{2}{3}A^2 \rangle \;.
\end{equation} 
Here the gauge field $A$ is regarded as a Lie-algebra-valued one form
and $\langle, \rangle $ represents metric in the field space that one obtains by construing a non-degenerate invariant
bilinear form on the Lie algebra space. $k$ is level for the theory and we  express $A= A^a_{\mu}\, T_a\, {\rm d}x^{\mu}$ where  $\{T_a\}$ are a particular basis of the
Lie-algebra. The equation of motion is given as,
\begin{equation}\label{eom} F \equiv d A + A \wedge A =0 . \end{equation}
For our purpose, the gauge group is ${\cal{N}}=(2,0)$ SuperPoincar\'{e} groups. The 3-manifold will be a one with a {\it boundary} and we shall identify the level $k$ with Newton's constant as $ k=\frac{1}{4 G}$. The basis elements $\{T_a\}$ are $J_a, P_a, Q_{\alpha}^{i}, T, Z$. Using the supertrace elements we get the corresponding supergravity action as,
\begin{eqnarray}\label{action2}
I_{\mu,,\bar \mu,\gamma}^{(2,0)} &= \frac{k}{4\pi} \int [2e^a \hat{R_a} + \mu L(\hat{\omega}_a) - \bar{\Psi}^i_{\beta} \nabla \Psi_{i}^{\beta} -2BdC + \bar \mu BdB] ,  
\nonumber \\
&A= e^a P_a + \hat{\omega}^a J_a + \psi_i^{\alpha} Q^i_{\alpha} + BT + CZ, 
\end{eqnarray}
where $\hat{\omega}^a = \omega^a +\gamma e^a$, for some constant $\gamma$ and $\bar{\Psi}^i_{\beta} $ is the Majorana conjugate gravitino. The ${\cal{N}}=2$ supergravity theory of \cite{Fuentealba:2017fck,Basu:2017aqn} is recovered in $\mu=\bar \mu = \gamma=0$ limit. The curvature two form $\hat R_a$, Lorentz Chern-Simons three form $L(\hat{\omega}_a)$ and the covariant derivative of the gravitino of \eqref{action2} can respectively be defined as,
\begin{align}\nonumber
\hat{R_a} =& d\hat{\omega}_a + \frac{1}{2} \epsilon_{abc}\hat{\omega}^b\hat{\omega}^c\\
L(\hat{\omega}_a) =& \hat{\omega^a} d\hat{\omega}_a + \frac{1}{3} \epsilon^{abc} \hat{\omega}_a \hat{\omega}_b\hat{\omega}_c\\
\nabla \Psi_{i}^{\beta} =& d \Psi_{i}^{\beta} + \frac{1}{2} \hat{\omega}^a \Psi_{i}^{\delta} (\Gamma^a)^{\beta}_{\delta} + B \Psi_{j}^{\beta} \epsilon^{ij}.\nonumber
\end{align}
Since CS theory is a gauge theory,
the equation of motion \eqref{eom} implies that locally the solutions of a CS field are pure gauge $A= G^{-1}d G,$ where $G$ is a local group element. Defining $\hat \omega= \frac{1}{2}\hat \omega^a \Gamma_a $, $e= \frac{1}{2}e^a \Gamma_a $ and $\mathcal{G}^1 = \frac{1}{2}(\Psi_1 -i \Psi_2)$ and $\mathcal{G}^2 = \frac{1}{2}(\Psi_1 + i \Psi_2)$  the onshell configuration for various fields of \eqref{action2} can be written as
\begin{align}\label{wb}
\hat{\omega} &= \Lambda^{-1}d\Lambda~~~~~~~~~~
B= d\tilde{B}. \nonumber \\
\mathcal{G}_1& = e^{-i\tilde{B}} \Lambda^{-1} d\eta_1, \quad 
\mathcal{G}_2 = e^{i\tilde{B}} \Lambda^{-1} d\eta_2. \nonumber \\
C &= -i (\bar{\eta}_{1\alpha} d\eta_2^{\alpha} - \bar{\eta}_{2\alpha} d\eta_1^{\alpha} + d\tilde{C} ) \nonumber \\
e &= - \Lambda^{-1} [\frac{1}{2} (\eta_1 \bar{d\eta_2} - \frac{1}{2} \eta_1 d\bar{\eta_2} \textbf{I})+ \frac{1}{2} (\eta_2 \bar{d\eta_1} - \frac{1}{2}\eta_2 d\bar{\eta_1} \textbf{I}) + db] \Lambda.
\end{align}
Here $\Lambda$ is an arbitrary $SL(2,R)$ group element of unit determinant and $\Gamma_a$ are generator of $SL(2,R)$. $B, C$ are $SL(2,R)$ scalars, $\eta_i, i=1,2$ are Grassmann-valued $SL(2,R)$ spinors and $b$ is a traceless $2 \times 2$ matrix. All these are local functions of three space time coordinates $u, \phi, r$. Since we are dealing with a gauge theory, we further choose a (radial) gauge condition $\partial_{\phi}A_r=0$ and hence the group element  splits as $G(u,\phi,r)= g(u,\phi)h (u,r)$.  Thus the gauge fields have following form, $$A=h^{-1}(a+d)h,  \qquad a= g^{-1}dg= a_u (u,\phi) du + a_{\phi} (u,\phi)d \phi.$$  We further choose that asymptotically $h= e^{-r P_0}$ and hence $\dot h(u,r)=\frac{\partial h(u,r)}{\partial u}=0$ at the boundary.
Advantage of this gauge choice is that the dependence in the radial coordinate
is completely absorbed by the group element $h$. Hence the boundary can be assumed to be uniquely 
located at any arbitrary fixed value of $r = r_0$, in particular to infinity. Thus the boundary  describes a two-dimensional
timelike surface with the topology of a cylinder . The radial gauge condition makes the above solutions of various field parameters  decomposed as,
\begin{eqnarray}\label{gff}
	\Lambda &=& \lambda(u,\phi) \zeta(u,r)  \nonumber \\ 
\tilde {B} &=& a(u,\phi) + \tilde{a}(u,r), \quad	\tilde{C} = c(u,\phi) + \tilde{c}(u,r) + \bar{d_2} \lambda \tilde{d_1} -  \bar{d_1} \lambda \tilde{d_2}  \nonumber \\
	\eta_1 &=& e^{i a}(\lambda\tilde{d}_1 (u,r)+ d_1(u,\phi) ), \quad 	\eta_2 = e^{-i a}(\lambda \tilde{d}_2 (u,r)+ d_2(u,\phi))\\
	b &=& \lambda E(u,r) \lambda^{-1} -\frac{1}{2} (d_1 \bar{\tilde{d_2}} \lambda^{-1} - d_1 \bar{\tilde{d_2}} \lambda^{-1} \textbf{I})- \frac{1}{2} (d_2 \bar{\tilde{d_1}} \lambda^{-1} - d_2 \bar{\tilde{d_1}} \lambda^{-1} \textbf{I}) + F(u,\phi), \nonumber
\end{eqnarray}
where $\dot  \zeta(u,r_0) = \dot {\tilde{a}}(u,r_0) = \dot {\tilde{c}}(u,r_0)= \dot {\tilde{d}}_1(u,r_0)= \dot {\tilde{d}}_2(u,r_0)= \dot E(u,r_0)=0.$ Thus even onshell, the system contains arbitrary local functions $ \lambda,F,a, c,d_1,d_2$ of time $u$ (and $\phi$) as residual degrees of freedom of the gauge system. 

In \cite{Fuentealba:2017fck} for ${\cal{N}}=2$ supergravity, the asymptotic fall of condition on the $r-$independent part of the gauge field was given as,
\begin{align}\label{bgf}
a = &\sqrt{2}[ J_1 + \frac{\pi}{k} (\mathcal{P}-\frac{4\pi}{k} \mathcal{Z}^2)J_0 + \frac{\pi}{k} (\mathcal{J} + \frac{2\pi}{k} \tau \mathcal{Z}) P_0 -\frac{\pi}{k} \psi_i Q^i_{+}  -\frac{2\pi}{k}\mathcal{Z} T - \frac{2\pi}{k} \tau Z]d\phi\\ \nonumber
& +[\sqrt{2}P_1 + \frac{8\pi}{k}\mathcal{Z}Z+ \frac{\pi}{k}(\mathcal{P}-\frac{4\pi}{k}\mathcal{Z}^2)P_0]du,
\end{align}
Here $\mathcal{P}, \mathcal{J},\mathcal{Z}, \tau, \psi_i$ are functions of $u,\phi$ only. These are the residual degree of freedoms that correspondence to $ \lambda,F,a, c,d_1,d_2$ as introduced above in \eqref{gff}. We do not consider the holonomy terms and hence the resulting action principle at the boundary only captures the asymptotic symmetries of the original gravitational theory.

\section{${\cal{N}}=2$ SuperPoincar\'{e} Wess-Zumino-Witten model and its Symmetries}\label{sec3}
In this section, we shall write down the dual WZW model that describes the dynamics of the above theory \eqref{action2}. For this purpose, notice that 
the asymptotic gauge field \eqref{bgf} is highly constrained. Firstly  its $u$ and $\phi$ components are related as, 
\begin{align}\label{bc1}
e^a_u = \omega^a_{\phi}, ~~~~~~~\omega^a_u =0, ~~~~~~~~~\psi^{\pm}_{Iu}=0, \qquad
B_u = 0,~~~~~~-4B_{\phi} = C_u.
\end{align}
The $u$ component of the gauge field \eqref{bgf} is further constrained as,
\begin{align}\label{bc2}
\hat\omega^1_{\phi}=\sqrt{2}; \quad \omega^2_{\phi}= 0; \quad \psi^{1+}_{\phi}= \psi^{2+}_{\phi} = 0; \quad e^1_{\phi}=e^2_{\phi}=0.
\end{align}
As the gauge field does not vanish at the boundary, for a well defined variational principle, we need to  the surface term to the action. At the boundary the surface term looks like:\\
\begin{equation}
I_{surf} =  - \frac{k}{2 \pi}\int du \tilde d\langle A_u, \delta \tilde A \rangle_{r_0 \rightarrow \infty}=-\frac{k}{4\pi} \int_{\partial M} du d\phi [\omega^a_\phi \omega_{a\phi} + 4 B^{2}_{\phi}] _{r_0 \rightarrow \infty},
\end{equation}
where $\phi-$ total derivative has been set to zero as $\phi$ is a compact direction.
Using the field parameters as defined in \eqref{action2}, the supertrace elements of the algebra  and the configuration \eqref{gff}, the  total onshell action  can be expressed as:
\begin{align}\label{sa}
I_{(2,0)}
=&\frac{k}{4\pi}\bigg\{ \int du d\phi Tr[2\mu \lambda^{-1} \lambda' \lambda^{-1} \dot{\lambda} -2( {\bar{d_1}}' \dot{d_2}+{\bar{d_2}}' \dot{d_1}) - 2i {a}' (\bar{d}_{1} \dot{d_2} - \bar{d}_{2}\dot{d_1} + \dot{\lambda} \lambda^{-1}(d_2 \bar{d_1} - d_1 \bar{d_2}) ) \nonumber\\ &- 4 (a')^2 
 -4 \dot{\lambda} \lambda^{-1} (\frac{1}{2} (d_1 \bar{d}'_2 - d_1 \bar{d_2}' \textbf{I}) + \frac{1}{2} (d_2 \bar{d}'_1 - d_2 \bar{d_1}' \textbf{I}) + F') - 2 (\lambda^{-1} \lambda')^2+ 2i \dot{a} c' + \bar{\mu} a' \dot{a}]\nonumber\\&+ \frac{2\mu}{3} \int Tr[(d\Lambda \Lambda^{-1})^3]\bigg\},
\end{align}
One can convince himself that the above action describes  a chiral WZW model with gauge group $SL(2,R)$.
We refer the readers to \cite{Banerjee:2019lrv} for the details computations required to arrive at the above result. The system shows gauge invariance under following (infinitesimal)gauge transformation
\begin{equation}\label{GT}
\begin{aligned}
\delta\lambda&=\beta\lambda, \quad
\delta d_i&=\beta d_i, \quad
\delta F&=[\beta,F],
\end{aligned}
\end{equation}
where the transformation parameter 
\begin{equation}
\beta=\left( \begin{array}{cc}{\beta_1} & {\beta_2} \\ {\beta_3} & {\beta_4}\end{array}\right),
\end{equation}\\
is a function of $u$ and other fields do not transform. 

\subsection{Global Symmetries of The Chiral WZW Model}\label{sec3.1}

The WZW model of \eqref{sa}
 is invariant under a set of global symmetries. As shown in \cite{Banerjee:2019lrv}, various fields change in a coordinate $u,\phi$ dependent transformation under these symmetries  as,  
\begin{align} \label{gs}
a &\rightarrow a + A(\phi); \quad c\rightarrow c- 4iu A'; \quad
d_1 \rightarrow e^{-iA} d_1; \quad d_2 \rightarrow e^{iA} d_2 \nonumber\\
c &\rightarrow c + \mathcal{C}(\phi) \nonumber\\
\lambda &\rightarrow \lambda \theta^{-1}(\phi); \quad F \rightarrow F + u \lambda ( \theta^{-1}\theta') \lambda^{-1} \nonumber\\
F &\rightarrow F + \lambda N(\phi) \lambda^{-1}\\
d_1 &\rightarrow d_1 + \lambda D_1(\phi); \quad c \rightarrow c + \bar{D_1}(\phi) \lambda^{-1} d_2; \quad F \rightarrow F -\frac{1}{2} d_2 \bar{D_1}(\phi) \lambda^{-1} \nonumber\\
d_2 &\rightarrow d_2 + \lambda D_2(\phi); \quad c \rightarrow c - \bar{D_2}(\phi) \lambda^{-1} d_1; \quad F \rightarrow F - \frac{1}{2} d_1 \bar{D_2}(\phi) \lambda^{-1} \nonumber
\end{align}
In each of the above expressions, the fields that are not written remain unchanged under that corresponding symmetry transformation. Symmetries are generated by scalar parameters $A(\phi),\mathcal{C}(\phi)$, matrix valued parameters $\theta(\phi),N(\phi)$ and spinor parameters $D_1(\phi),D_2(\phi)$. These parameters are independent of $u$ and thus they represent Global symmetry transformations. 

The conserved currents corresponding to the above symmetries have also been constructed in \cite{Banerjee:2019lrv}. Below we present those currents,
\begin{align}\label{cc}
J^{\mu}_{A} &=\delta^{\mu}_{0} \frac{k}{4\pi}Tr [2 \bar{\mu} a' + 2i c' - 8u a''+2i(\bar{d_2}' d_1 - \bar{d_1}' d_2 - ia' (\bar{d_2}d_1 + \bar{d_1} d_2))]A =\delta^{\mu}_{0}[(-Q^A) (-A)]\nonumber\\
J^{\mu}_{C} &= \delta^{\mu}_{0}\frac{k}{4\pi}Tr[2i a' \mathcal{C}] = \delta^{\mu}_{0} [Q_C (-i \mathcal{C})], \quad Q_C = -\frac{k a'}{2 \pi}\nonumber\\
J^{\mu}_{\Theta} &= \delta^{\mu}_{0}\frac{k}{2\pi} Tr[\{\lambda^{-1}\hat\alpha \lambda + 2u (\lambda^{-1}\lambda')'- 2\mu \lambda^{-1}\lambda' \}\Theta]=  \delta^{\mu}_{0} 2 Tr[Q^J_a \Theta^a]\nonumber\\
J^{\mu}_{N} &= \delta^{\mu}_{0} \frac{k}{4\pi}Tr[-4\lambda^{-1}\lambda' N] = \delta^{\mu}_{0} 2 Tr[Q^P_a (-N^a)]\\
J^{\mu}_{D_2} &= \delta^{\mu}_{0}(-\frac{k}{\pi})Tr [(\bar{d_1}' \lambda + i a' \bar{d_1} \lambda) D_2] = \delta^{\mu}_{0}Tr[Q^{G_2}_{\alpha}D_2^{\alpha}]\nonumber\\
J^{\mu}_{D_1} &= \delta^{\mu}_{0}(-\frac{k}{\pi}) Tr[(\bar{d_2}' \lambda - i a' \bar{d_2} \lambda) D_1]= \delta^{\mu}_{0}Tr[Q^{G_1}_{\alpha}D_1^{\alpha}],\nonumber
\end{align}
where $N(\phi)$ and $\Theta(\phi)$ are infinitesimal $SL(2,\mathbb{R})$ matrices which can be further expanded in the basis of $\Gamma$ matrices as $N(\phi) = N^a(\phi) \Gamma_a$ and $\Theta(\phi) = \Theta^a(\phi) \Gamma_a$. Other parameters have also been considered as infinitesimal.  It can be checked that the above currents satisfy the following current algebra,
\begin{align}\label{CA}
\left\lbrace  Q^P_a(\phi), Q^P_b(\phi')    \right\rbrace_{DB} &= \left\lbrace  Q^P_a(\phi), Q^A(\phi')    \right\rbrace_{DB}=\left\lbrace   Q^P_a(\phi), Q^C(\phi')    \right\rbrace_{DB}=0  \nonumber\\
\left\lbrace  Q^P_a(\phi),  Q^{G_1}_{\alpha}(\phi')    \right\rbrace_{DB} &=\left\lbrace  Q^P_a(\phi), Q^{G_2}_{\alpha}(\phi')    \right\rbrace_{DB} =0 \nonumber\\
\left\lbrace  Q^P_a(\phi), Q^J_b(\phi')    \right\rbrace_{DB} &= \left\lbrace  Q^J_a(\phi), Q^P_b(\phi')    \right\rbrace_{DB} =\epsilon_{abc} Q^P_c(\phi) \delta(\phi - \phi')- \frac{k}{2\pi} \eta_{ab} \partial_{\phi}\delta(\phi-\phi')\nonumber\\
\left\lbrace  Q^J_a(\phi), Q^J_b(\phi')    \right\rbrace_{DB} &= \epsilon_{abc} Q^J_c(\phi) \delta(\phi - \phi')+\mu \frac{k}{2\pi} \eta_{ab} \partial_{\phi}\delta(\phi-\phi')\nonumber\\
\left\lbrace  Q^J_a(\phi), Q^A(\phi')    \right\rbrace_{DB} &=\left\lbrace   Q^J_a(\phi), Q^C(\phi')    \right\rbrace_{DB}=0\\
\left\lbrace  Q^{G_1}_{\alpha}(\phi), Q^J_a(\phi')    \right\rbrace_{DB} &=-\frac{1}{2} (\Gamma_a)^{\beta}_{\alpha} Q^{G_1}_{\beta}(\phi) \delta(\phi-\phi')\nonumber\\
\left\lbrace  Q^{G_2}_{\alpha}(\phi), Q^J_a(\phi')    \right\rbrace_{DB} &=-\frac{1}{2} (\Gamma_a)^{\beta}_{\alpha} Q^{G_2}_{\beta}(\phi) \delta(\phi-\phi')\nonumber\\
\left\lbrace  Q^C(\phi), Q^C(\phi')    \right\rbrace_{DB} &=\left\lbrace  Q^C(\phi),  Q^{G_1}_{\alpha}(\phi')    \right\rbrace_{DB} =\left\lbrace  Q^C(\phi), Q^{G_2}_{\alpha}(\phi')    \right\rbrace_{DB} =0\nonumber\\
\left\lbrace  Q^C(\phi), Q^A(\phi')    \right\rbrace_{DB} &= \frac{k}{2\pi } \partial_{\phi} \delta(\phi-\phi'),\quad
\left\lbrace  Q^{A}(\phi), Q^{A}(\phi')    \right\rbrace_{DB} = \frac{k}{2\pi} \bar{\mu} \partial_{\phi} \delta(\phi-\phi')\nonumber\\
\left\lbrace  Q^{G_1}_{\alpha}(\phi), Q^{A}(\phi')    \right\rbrace_{DB}  &=- i Q^{G_1}_{\alpha}(\phi) \delta(\phi - \phi'), \quad
\left\lbrace  Q^{G_2}_{\alpha}(\phi), Q^{A}(\phi')    \right\rbrace_{DB} = i Q^{G_2}_{\alpha}(\phi) \delta(\phi - \phi')\nonumber\\
\left\lbrace  Q^{G_1}_{\alpha}(\phi), Q^{G_2}_{\beta}(\phi')   \right\rbrace_{DB} &= -(C\Gamma^a)_{\alpha\beta} Q^P_a \delta(\phi-\phi') - \frac{k}{\pi} C_{\alpha \beta} \partial_{\phi} \delta(\phi-\phi') +ia' \frac{k}{\pi} C_{\alpha\beta} \delta(\phi-\phi'),\nonumber
\end{align}
Next, we notice that constraints of \eqref{bc2} further implies that the canonical current generators are constrained as,
\begin{align}\label{2cc}
Q^P_0 &= \sqrt{2} \frac{k}{2\pi}, \quad Q^J_0 = -\sqrt{2} \frac{\mu k}{2\pi} \nonumber\\
Q^1_{+}&=0, \qquad Q^2_{+}=0\\
Q^P_2 &=0, \quad Q^J_2 =0.\nonumber
\end{align} 
The above relations can be expressed as constraints on various fields of the WZW model \eqref{sa}. Thus we see that when the asymptotic boundary condition of \eqref{bgf} is explored to its full capacity, the theory gets more constrained. The first four of \eqref{2cc} are first class constraints and they will produce a gauge invariance for the system. To understand the proper symmetry structure of the theory, we shall need gauge invariant canonical symmetry generators.
Hence we implement a modified Sugawara construction to define these gauge invariant currents as, 
\begin{align}\label{bms3G}
{\cal{H}} = H + \partial_{\phi} Q^P_2; \hspace{15pt} {\cal{J}} = J - \partial_{\phi} Q^J_2; \hspace{15pt} \hat{\mathcal{G}}^I= \mathcal{G}^I + \partial_{\phi} Q^I_{+}, Q^A, Q^C,
\end{align}
where we have defined 
\begin{align}\label{ag}
H &= \frac{\pi}{k} Q^P_a Q^P_a +4 \frac{\pi}{k} Q^C Q^C \nonumber\\
J &= -\mu \frac{\pi}{k} Q^P_a Q^P_a - 2 \frac{\pi}{k} Q^J_a Q^P_a + \frac{\pi}{k} C_{\alpha \beta} Q^{G_1}_{\alpha} Q^{G_2}_{\beta}+ 2 \frac{\pi}{k} Q^A Q^C- \bar{\mu} \frac{\pi}{k} Q^C Q^C\\
\mathcal{G}^1 &= \frac{\pi}{k} (Q^P_2 Q^{G_1}_+ +\sqrt{2} Q^P_0 Q^{G_1}_-)+2i \frac{\pi}{k}  Q^{G_1}_+ Q^C, \nonumber \\
\mathcal{G}^2 &=\frac{\pi}{k} (Q^P_2 Q^{G_2}_+ +\sqrt{2} Q^P_0 Q^{G_2}_-)- 2 i \frac{\pi}{k}  Q^{G_1}_+ Q^C \nonumber.
\end{align}
It is easy to see that the above conserved charges close to following  generalised quantum  superBMS$_3$ algebra  on the constrained surface  as,
\begin{align}\label{n20pb}
\lbrace {\cal{J}}(\phi),  {\cal{J}}(\phi') \rbrace _{DB}&= ( {\cal{J}}(\phi) +  {\cal{J}}(\phi')) \partial_{\phi} \delta(\phi - \phi') -\mu  \frac{k}{2\pi} \partial_{\phi}^3 \delta(\phi-\phi')\nonumber\\
\lbrace {\cal{H}}(\phi),  {\cal{J}}(\phi') \rbrace_{DB} &= ({\cal{H}}(\phi) + {\cal{H}}(\phi')) \partial_{\phi} \delta(\phi - \phi')- \frac{k}{2\pi} \partial_{\phi}^3 \delta(\phi-\phi')\nonumber\\
\{\tilde{\mathcal{H}}(\phi),\tilde{\mathcal{H}}(\phi')\}_{DB}&=0, \quad \lbrace{\cal{H}}(\phi), Q^A(\phi') \rbrace_{DB} =4 Q^C(\phi) \partial_{\phi} \delta(\phi - \phi')\nonumber\\
\lbrace  {\cal{J}}(\phi), Q^A(\phi') \rbrace_{DB} &=  Q^A(\phi) \partial_{\phi}\delta(\phi-\phi'), \quad \lbrace  {\cal{J}}(\phi), Q^C(\phi') \rbrace_{DB} = Q^C(\phi)\partial_{\phi} \delta(\phi - \phi')\nonumber\\
\lbrace  {\cal{J}}(\phi), Q^A(\phi') \rbrace_{DB} &=  Q^A(\phi) \partial_{\phi}\delta(\phi-\phi')\\
\left\lbrace  Q^C(\phi), Q_A(\phi')    \right\rbrace_{DB} &=  \frac{k}{2\pi} \partial_{\phi} \delta(\phi-\phi'), \quad \left\lbrace  Q^A(\phi), Q^A(\phi')    \right\rbrace_{DB} =  \frac{k}{2\pi} \bar \mu\partial_{\phi} \delta(\phi-\phi')\nonumber\\
\lbrace  {\cal{J}}(\phi), \hat {\mathcal{G}}^i(\phi') \rbrace_{DB} &= (\hat {\mathcal{G}}^i(\phi) +\frac{1}{2} \hat {\mathcal{G}}^i(\phi'))\partial_{\phi} \delta(\phi - \phi'), \quad (i=1,2)\nonumber\\
\lbrace  {\cal{H}}(\phi), \hat {\mathcal{G}}^i(\phi') \rbrace_{DB} &= 0, \quad (i=1,2)\nonumber\\
\lbrace  \hat {\mathcal{G}}^1(\phi),Q^A(\phi')  \rbrace_{DB} &= -i  \hat {\mathcal{G}}^1(\phi) \delta (\phi-\phi'), \quad
\lbrace \hat {\mathcal{G}}^2(\phi),Q^A(\phi')  \rbrace_{DB} = i \hat {\mathcal{G}}^2(\phi) \delta (\phi-\phi')\nonumber\\
\lbrace \hat {\mathcal{G}}^1(\phi),\hat {\mathcal{G}}^2(\phi')  \rbrace_{DB} &=  {\cal{H}}(\phi)\delta(\phi-\phi')-\frac{k}{\pi}\partial^2_{\phi}\delta(\phi-\phi') -2i(Q^C(\phi)+Q^C(\phi'))\delta'(\phi-\phi') .\nonumber
\end{align}

Let us next gauge the symmetries introduced in the beginning of subsection \ref{sec3.1} in \eqref{gs}. As we have already mentioned, there are four first class constraints as noted in \eqref{2cc} and they will produce four gauge symmetries to the system. Thus it is clear to see that last four transformations of \eqref{gs} can be gauged, i.e. the transformation parameters can be made a local function of $u$ as well. Below we present the gauged version of the chiral WZW model \eqref{sa}.

\subsection{Gauging the Chiral WZW model}\label{sec3.2}
Imposing the first class constraints of \eqref{bc2} on onshell gauge field parameters imply following relations  
\begin{align*}
(\lambda^{-1}\lambda')^1 = \sqrt{2} \qquad (\lambda^{-1}\frac{\hat\alpha}{2}\lambda)^1=0\\
(\lambda^{-1} d'_1 + ia' \lambda^{-1} d_1)^- =
(\lambda^{-1} d'_2 - ia' \lambda^{-1} d_2)^- =0.
\end{align*}
The above relations can be equivalently re-caste in terms of global symmetry currents as given in \eqref{2cc}.  Here we are setting a part of currents to constant or zero value that comes from symmetry transformations involving bosonic symmetry transformation parameter $N,\Theta$ along $\Gamma_0$ and fermionic parameters $[\bar{D_1}]_+=[\bar{D_2}]_+=0$. To gauge the corresponding symmetries, one needs four "gauge fields" corresponding to four constrained currents $J$. Since  the currents are nontrivial  along $u$ directions, only the $u-$component of the gauge fields will appear in the modified action. In general, for gauging a global symmetry, we need to replace the ordinary derivatives on various fields by the corresponding covariant ones. For WZW model on a Lie group $G$, only special subgroups of $G$ can be gauged, as otherwise, the WZW term makes it anomalous. The detailed procedure of gauging WZW model has been greatly described in a seminal paper \cite{FEHER19921} where it has been noted that only subgroups  generated by root vectors associated with positive and negative roots can be gauged. This implies the subgroup elements must be nilpotent matrices. Similar strategy has already been implemented in \cite{Barnich:2013yka,Barnich:2015sca} for gauging  $SL(2,R)$ chiral WZW models with(out) minimal supersymmetric extensions. We shall follow the procedure of \cite{Henneaux:1999ib} which is also similar in spirit. In this case, we introduce four Lagrange multipliers for gauging the constrained (only first class ones) currents. Since the constrained currents are along $\Gamma_0$, the above criteria is satisfied. Using these multipliers we write down an improved action, where the improvement term is local. Further,  the transformations of the Lagrange multipliers are derived by demanding that the full improved action is invariant under the above mentioned gauging of symmetries. The improved action looks like,
\begin{equation}\label{gsa}
 I[\lambda,c,a,F,d_1,d_2,\Psi,A_\mu]=I[\lambda,c,a,F,d_1,d_2]+I_g
\end{equation}
where, $I[\lambda,C,F,d_1,d_2]$ is as given in \eqref{sa} and
\begin{equation}
\begin{aligned}
I_g&=\frac{k}{\pi}\int dud\phi\operatorname{Tr}\Bigg[A_u(\lambda^{-1}\frac{\hat\alpha}{2}\lambda)+\Tilde{A_u}\lambda^{-1}\lambda^{\prime}-\mu_M\Tilde{A_{u}}\\&+\frac{1}{2}[\lambda^{-1}(d_1^\prime+ia^\prime d_1)]\Bar{\Psi}_2+\frac{1}{2}[\lambda^{-1}(d_2^\prime-ia^\prime d_2)]\Bar{\Psi}_1\Bigg]
\end{aligned}
\end{equation}
Here, $I_g$ is a local function of "gauge fields" $A, \tilde A , \bar \Psi_i , \quad i=1,2$ . 
$A_u$, $\Tilde{A_u}$ are along $\Gamma_0$ and $[\Bar{\Psi}_1]_+=0=[\Bar{\Psi}_2]_+$. Further  we have chosen $\mu_M:=\tilde \mu\Gamma_1$\footnote{$\Gamma_1$ is also nilpotent matrix.}, where $\tilde \mu$ is an arbitrary constant to be able to set the currents to required constant value.
It can indeed be checked that the above modified action \eqref{gsa} in invariant under following four gauge transformations,
\begin{align}
 T_1:& \quad \delta_NF =\lambda N\lambda^{-1}, \quad  \delta_N\lambda=\delta_Nd_1=\delta_Nd_2=\delta_Nc=\delta_Na=0  \\
&\delta_N\Psi_1=\delta_N\Psi_2=\delta_N A_u=0, \quad
\delta_N\Tilde{A_u}=(\dot{N}+[A_u,N]) \nonumber\\
 T_2:& \quad \delta_\Theta\lambda =-\lambda\Theta, \quad \delta_\Theta F=u\lambda\Theta^{-1}\Theta^\prime\lambda^{-1}, \quad \delta_\Theta d_1=\delta_\Theta d_2=\delta_Nc=\delta_Na=0
\\
&\delta_\Theta A_u=-(\dot{\Theta}+[A_u,\Theta]), \quad
\delta_\Theta\Psi_1=\delta_\Theta\Psi_2=0, \quad
\delta_\Theta\Tilde{A_u}=u(\dot{\Theta}^\prime+[A_u,\Theta^\prime])+ \frac{\mu}{2}\dot\Theta-[\Tilde{A_u},\Theta] \nonumber\\
 T_3:& \quad \delta_{D_1}\lambda =\delta_{D_1}d_2=\delta_Na=0, \quad \delta_{D_1}d_1=\lambda D_1, \quad 
\delta_{D_1}C=\Bar{D_1}\lambda^{-1}d_2, \quad
\delta_{D_1}F=-\frac{1}{2}d_2\Bar{D_1}\lambda^{-1} \nonumber\\
&\delta_{D_1}A_u=\delta_{D_1}\Tilde{A_u}=\delta_{D_1}\Psi_2=0, \quad
\delta_{D_1}\Psi_1=-\partial_\mu\Bar{D_1}\\
 T_4:& \quad \delta_{D_2}\lambda=\delta_{D_2}d_1=\delta_Na=0 , \quad \delta_{D_2}d_2=\lambda D_2, \quad
\delta_{D_2}C=-\Bar{D_2}\lambda^{-1}d_1, \quad
\delta_{D_2}F=-\frac{1}{2}d_1\Bar{D_2}\lambda^{-1} \nonumber\\
&\delta_{D_2}A_u =\delta_{D_2}\Tilde{A_u}=
\delta_{D_2}\Psi_1=0, \quad
\delta{D_2}\Psi_2=-\partial_{\mu}\Bar{D_2}.
\end{align}
Note that all the parameters of the transformation mentioned here, depend on both $(u,\phi)$ and as said earlier,  $\Theta, N$ are along $\Gamma_0$, and $[\Bar{D_1}]_+=[\Bar{D_2}]_+=0$. 
The equations of motion of  the four non-dynamical Lagrange multiplier  $A_u,\Tilde{A_u}, \Bar{\Psi_1}, \Bar{\Psi_2}$ rightly reproduces back the constrained  relations as given in the beginning of this section if one chooses $\tilde\mu= \frac{1}{\sqrt 2}.$ Thus we conclude that \eqref{gsa} represents the gauged version of chiral WZW model \eqref{sa}, where we have gauged the specific part of global symmetries whose corresponding currents gives first class constraints. The gauge symmetry along $\Gamma_2$ are still present. In the next section, we shall write down the reduced phase space description for the WZW model \eqref{sa} and show that reduced action is an equivalent description of the above gauged chiral WZW model of \eqref{gsa}. We shall also comment on the equivalence of the residual symmetries of the two descriptions.

\section{Liouville Like theory}\label{sec4}
In this section, we present the reduced phase space description of the chiral WZW model of \eqref{sa}. For this purpose a particular decomposition of the fields,  known as Gauss Decomposition\footnote{Important aspects of Gauss decomposition are discussed in \cite{Tsutsui:1994pp}}, is useful.  The procedure is to expand the fields in the Chevalley-Serre basis of the corresponding gauge group, i.e. $SL(2,R)$ for the present case. Our conventions are listed in appendix \ref{AppA}. The decomposition is,
\begin{equation}
\lambda=e^{\sigma \Gamma_{1} / 2} e^{-\varphi \Gamma_{2} / 2} e^{\tau \Gamma_{0}} \quad, \quad F=-\bigg(\frac{\eta}{2} \Gamma_{0}+\frac{\theta}{2} \Gamma_{2}+\frac{\zeta}{2} \Gamma_{1}\bigg),
\end{equation}
where $\sigma, \varphi,\tau,\eta,\theta,\xi$ are scalar fields and are functions of both $u, \phi$. The Gaussian decomposition is useful as in this decomposition the 3D bulk part of  the WZW model \eqref{sa} simplifies to a total derivative term as\footnote{here we have assumed same notation for component fields but they have dependence on all three directions, $r,u,\phi.$},
$$\frac{2}{3}\operatorname{Tr}[(d\Lambda\Lambda^{-1})^3]=drdud\phi~~\epsilon^{\nu\gamma\delta} \partial_{\nu}\left(e^{-{\varphi}}~~\partial_{\gamma} \tau~~\partial_{\delta} \sigma\right).$$
Thus, using Stoke's formula, the bulk term can be reduced to a 2-dimensional integral.  This makes further computations technically simple. Two product operators that are mostly used are given as,
\begin{equation}
 \lambda^{-1}\lambda^\prime= \left( \begin{array}{cc}{{ -\sigma^\prime\tau e^{-\phi}-{\phi^\prime/2}}} & {-\sqrt{2}\sigma^\prime\tau^2 e^{-\phi}+\sqrt{2}\tau^\prime-\sqrt{2}\tau\phi^\prime} \\ {\frac{\sigma^\prime}{\sqrt{2}}e^{-\phi}} &{ \tau\sigma^\prime e^{-\phi}+\phi^{\prime}/2}\end{array}\right)    
\end{equation}
and
\begin{equation}
\dot{\lambda} \lambda^{-1}= \left( \begin{array}{cc}{-\frac{\dot{\phi}}{2}-\sigma \dot{\tau}e^{-\phi}} & {\sqrt{2}\dot{\tau} e^{-\phi}} \\ {\frac{\dot{\sigma}}{\sqrt{2}}-\frac{\sigma}{\sqrt{2}}\dot{\phi}-\frac{\sigma^2}{\sqrt{2}}\dot{\tau}e^{-\phi}} & {\frac{\dot{\phi}}{2}+\sigma \dot{\tau}e^{-\phi}}\end{array}\right) 
\end{equation}
The first class constraints can also be recast in terms of these newly defined fields . Let us first look at $Q^P_0$ condition. It reduces as,
\begin{equation}\label{1}
Q^P_0=\frac{\sqrt{2}k}{2\pi}\implies\sigma^\prime=\sqrt{2}e^{\varphi} .   
\end{equation}
Next we look at two fermionic current constraints. They are given as,
\begin{eqnarray}
Q^1_+=0 \implies&\bigg[-{d_1^-}^\prime+{d_1^+}^\prime\frac{\sigma}{\sqrt{2}}-{d_2^-}^\prime+\frac{\sigma}{\sqrt{2}}{d_2^+}^\prime\Big]+(ia^\prime)\Big[-{d_1^-}+{d_1^+}\frac{\sigma}{\sqrt{2}}+{d_2^-}-\frac{\sigma}{\sqrt{2}}{d_2^+}\bigg]=0 \nonumber \\
Q^2_+=0 \implies&\Big[{d_1^-}^\prime-{d_1^+}^\prime\frac{\sigma}{\sqrt{2}}-{d_2^-}^\prime+\frac{\sigma}{\sqrt{2}}{d_2^+}^\prime\Big]+(ia^\prime)\Big[{d_1^-}-{d_1^+}\frac{\sigma}{\sqrt{2}}+{d_2^-}-\frac{\sigma}{\sqrt{2}}{d_2^+}\Big]=0
\end{eqnarray}
Redefining new fermionic parameters as $d_1=e^{ia}d_1^N$ and $ d_2=e^{-ia}d_2^N$, the above two conditions can be written in a compact form as,
\begin{equation}\label{2}
{d_1^{N-}}^\prime=\frac{\sigma}{\sqrt{2}}{d_1^{N+}}^\prime  , \quad  
{d_2^{N-}}^\prime=\frac{\sigma}{\sqrt{2}}{d_2^{N+}}^\prime   .   
\end{equation}
Finally, we look at the $Q^J_0$ constraint. The reduced constraint looks like, 
\begin{equation*}
\begin{aligned}
Q^J_0=-\frac{\sqrt{2}\mu k}{2\pi}\implies&2\mu(\sqrt{2} e^{\varphi}-\sigma^\prime)+2u(\sigma^{\prime\prime}-\varphi^\prime\sigma^\prime )-\eta^\prime\sigma^2-2\sigma\theta^\prime +2\zeta^\prime \\&+(ia^\prime)[2\sigma d_2^+d_1^--2\sqrt{2}d_2^-d_1^--\sqrt{2}\sigma^2d_2^+d_1^++2\sigma d_2^-d_1^+]\\&+(\sigma )(d_2^+{d_1^-}^\prime+d_2^-{d_1^+}^\prime+d_1^+{d_2^-}^\prime+d_1^-{d_2^+}^\prime)+(\sqrt{2})(-d_2^-{d_1^-}^\prime-d_1^-{d_2^-}^\prime)\\&-(\frac{\sigma^2}{\sqrt{2}})(d_2^+{d_1^+}^\prime+d_1^+{d_2^+}^\prime)=0
\end{aligned}
\end{equation*}
Using the redefined fermions and the last three currents constraints, the above condition simplifies as,
\begin{equation}\label{3}
-\eta^\prime\sigma^2-2\sigma\theta^\prime+2\zeta^\prime=0    
\end{equation}
Equations \eqref{1},\eqref{2} and \eqref{3} represents the first class constraints in terms of the new fields.

\subsection{The Reduced Action}
We  present the action by computing various terms of \eqref{sa} in terms of above newly defined fields and reducing it further by using the constraints relations of the last section as\footnote{look at appendix \ref{AppB} for some details}
\begin{equation}\label{LA}
\begin{aligned}
I&=\frac{k}{4\pi}\int dud\phi\Bigg[\mu\varphi^\prime\dot{\varphi}+\xi^\prime\dot{\varphi}-{\varphi^\prime}^2- 2i\dot{a}^\prime D +\Bar{\mu}a^\prime\dot{a}-4(a^\prime)^2 + 2 \bigg( \dot \chi_1 \chi_2+\dot \chi_2 \chi_1\bigg) \Bigg],
\end{aligned}
\end{equation}
where the redefined fields are, $$\xi=2(\theta+\sigma\eta)+\bar d_2 d_1, \quad \chi_i= e^{\phi/2} \ d_i^{+}, \quad D= c + (d_2^{+}d_1^{-}-d_1^{+}d_2^{-}) - \sqrt{2}\sigma d_2^{+} d_1^{+}.$$ \eqref{LA} is a flat limit of a super-Liouville action with two supercharges and two internal $R-$symmetry fields. This is a generalised version of flat limit of super(Liouville actions presented in \cite{Barnich:2013yka,Barnich:2015sca}). To understand the connection with Liouville, we refer readers to appendix \ref{AppC}.  The above action \eqref{LA} is equivalent to the gauged chiral WZW model of \eqref{gsa}. Solving the algebraic equation of motions of the Lagrange multipliers and putting it back in \eqref{gsa} will exactly give us \eqref{LA}, when expressed in terms of Gauss Variables. 

Finally we present a realisation of superBMS$_3$ generators of \eqref{bms3G} in terms of Liouville fields. With a straight algebra and use of \eqref{1},\eqref{2},\eqref{3} they can be found as,
\begin{align}
\mathcal{H}&=\frac{k}{4\pi}\big[{\varphi^\prime}^2-2\varphi^{\prime\prime}+4(a^\prime)^2\big], \nonumber\\
\mathcal{J}=&\frac{k}{4 \pi}
\big[\xi^{\prime} \varphi^{\prime}-\xi^{\prime \prime}+u\big(2
 \varphi^{\prime \prime \prime}-2\varphi^{\prime} \varphi^{\prime \prime}-8a^{\prime\prime}a^\prime\big) +2(\chi_{1}^\prime \chi_{2}+\chi_{2}^\prime \chi_{1})\big] \nonumber\\  
& +2ia^\prime D^{\prime}+\bar{\mu}(a^{\prime})^2-12\mu(a^\prime)^2\big]+\mu \mathcal{H}, \nonumber\\
\hat{\mathcal{G}}^{1}&= \frac{k}{\pi}\Big((\frac{\varphi^{\prime}}{2}+\iota a^\prime)  \chi_2-{\chi_2}^{\prime}\Big), \quad
\hat{\mathcal{G}}^{2} = \frac{k}{\pi}\Big((\frac{\varphi^{\prime}}{2}-\iota a^\prime)  \chi_1-{\chi_1}^{\prime}\Big) \nonumber \\
Q^A&=-\frac{k}{4\pi}\operatorname{Tr}\big[2\Bar{\mu}a^\prime+2iD^\prime-8ua^{\prime\prime}+4i\chi_2\chi_1\big], \quad Q^C= - \frac{k a'}{ 2 \pi}.
\end{align}	
It can be  checked that the above generators constitute a set of global symmetries of the reduced action \eqref{LA}. To obtain the symmetry transformations, we first  find the canonical conjugate momenta of fields of \eqref{LA}. They are given as,
\begin{align}
{p_\varphi}=\frac{k}{4\pi}(\xi^\prime+\mu\varphi^\prime)~~~~~~~~~~~& p_\xi=\frac{k}{4\pi}\varphi^\prime & {p_a}=\frac{k}{4\pi}({\bar\mu}a^\prime+2iD^\prime) \nonumber \\
p_D=\frac{k}{4\pi}2ia^\prime~~~~~~~~~~~~~~& p_{\chi_1}=\frac{k}{4\pi}4\chi_2 & p_{\chi_2}=\frac{k}{4\pi} 4\chi_1
\end{align}
To obtain variation of various fields we use Hamiltonian formulation. The variation of fields can be computed from their Poisson brackets with the global charge as
\begin{equation}
-\delta A=\{A,Q\}    
\end{equation}
where,
\begin{equation}
Q=\int^{2\pi}_0d\phi\Bigg[\mathcal{H}T+\mathcal{J}Y+Q^AB+Q^CK+2\mathcal{\mathcal{G}}^1\epsilon_1+2\mathcal{\mathcal{G}}^2\epsilon_2\Bigg],
\end{equation}
and $T,Y,B,K,\epsilon_1,\epsilon_2$ are $\phi$ dependent symmetry transformation parameters. Transformation of various fields are then given as,
\begin{align}
-\delta\varphi&=Y\varphi^\prime+Y^\prime\\
-\delta\xi&=2f\varphi^\prime+\xi^\prime Y+2f^\prime-4\epsilon_1\chi_2-4\epsilon_2\chi_1\\
-\delta a&={a^\prime Y}{-B}\\
{-\delta D}&={D^\prime Y}-4ia^\prime T+4iuB^\prime+iK-4iua^\prime Y^\prime-4\epsilon_1\chi_2+4\epsilon_2\chi_1-ia^\prime\bar{\mu}Y+i\bar{\mu}B\\
{-\delta\chi_1}&=-\chi_1^\prime Y-\frac{1}{2}Y^\prime\chi_1
+2\Big(\frac{\varphi^\prime}{2}+ia^\prime\Big)\epsilon_1+2\epsilon^\prime_1{-i\chi_1B}\\
{-\delta\chi_2}&=-\chi_2^\prime Y-\frac{1}{2}\chi_2Y^\prime+2\Big(\frac{\varphi^\prime}{2}{-}ia^\prime\Big)\epsilon_2+2\epsilon^\prime_2{+i\chi_2B}
\end{align}
where,
\begin{equation}
f=T(\phi)+\mu Y(\phi)+uY^\prime  .  
\end{equation}
It can be checked that above transformations are global (u-independent)  symmetry of the reduced action \eqref{LA}, as expected. The algebra  of the corresponding Noether charges is again superBMS$_3$ of \eqref{n20pb}. The system is also invariant under transformations $$\delta \varphi= F_1(u),  \delta\xi = F_2(u), \delta D= F_3(u). $$ Let us briefly elaborate on the source of these local $u-$ dependent symmetries : the symmetry transformation of $\xi, D$ is an artefact of the form of the reduced action, as it involves only $\phi$ derivative of these fields. The symmetry transformation of $\varphi$ is related to the gauge invariance of \eqref{sa}, as given in \eqref{GT}.
For the Gauss decomposed fields, \eqref{GT} implies following transformation
\begin{equation}
\begin{aligned}
\delta\varphi&=-2\beta_1-\sqrt{2}\beta_2\sigma, \quad
\delta\theta=\sqrt{2}\beta_2\zeta-\sqrt{2}\beta_3\eta\\
\delta\eta&=(\beta_1-\beta_4)\eta-\sqrt{2}\beta_2\theta, \quad
\delta\sigma=\sqrt{2}\beta_3+\beta_4\sigma-\beta_1\sigma-\beta_2\frac{\sigma^2}{\sqrt{2}}\\
\delta d_i^+&=\beta_1d_i^++\beta_2d_i^-, \quad
\delta d_i^-=\beta_3d_i^++\beta_4d_i^-
\end{aligned}
\end{equation}
with $\beta_1 +\beta_4 = 0 $. Hence we can decompose $\beta$ matrix in the basis of $SL(2,R)$ generators, like field $F$. Here $\beta_1, \beta_2, \beta_3$ are three independent transformation parameters that depend on $u$. For the reduced fields of \eqref{LA}, we get
\begin{align}
\delta\varphi&=-2\beta_1-\sqrt{2}\beta_2\sigma, \quad \delta\xi 
=2\sqrt{2}\beta_2\zeta-2\sqrt{2}\sigma\beta_2\theta-\sqrt{2}\beta_2\eta\sigma^2 \\
\delta D & = 2 \beta_2 d_2^- d_1^- - \sqrt{2} \sigma \{ \beta_2 d_2^- d_1^+ + \beta_2 d_2^+ d_1^-\} + \beta_2 \sigma^2 d_2^+ d_1^+ \\
\delta\chi_i&=-\frac{\beta_2}{\sqrt{2}}\sigma e^{\varphi/2}d_i^++e^{\varphi/2}\beta_2d_i^-  .  
\end{align}
It can be checked that the above transformations, in presence of all three parameters $\beta_1, \beta_2, \beta_3$ are not symmetries of \eqref{LA}. Instead,  transforms of $\varphi$ by turning on only $\beta_1$ is a symmetry of the action. Turning on $\beta_1$ implies gauge transformation along $\Gamma_2$, which is the only non-nilpotent generator of $SL(2,R)$. This is the residual gauge symmetry of the gauged chiral WZW model of \eqref{gsa}. Thus we see that both the reduced phase space Lagrangian \eqref{LA} and the gauged chiral WZW model \eqref{gsa} preserves identical global and gauge symmetry.  

\section{Outlook}\label{sec5} 

In this paper, we have presented three equivalent descriptions of ${\cN=2}$ three dimensional supergravity theory. The first description in terms of a chiral WZW model was derived in \cite{Banerjee:2019lrv}, where as the other two equivalent descriptions in terms of a gauged version of the chiral WZW model and a flat limit of generalised super-Liouville theory have been derived in this paper. All these theories are invariant under most generic quantum ${\cN=2}$ superBMS$_3$ symmetry constructed in \cite{Banerjee:2019lrv} at null infinity.

One interesting point to note here is that Liouville theory can also be viewed as  free field theory under proper B$\ddot a$clund transformations \cite{Barnich:2012rz}. In \cite{Banerjee:2015kcx,Banerjee:2016nio} we have presented a free field realisation of BMS$_3$ algebra and its supersymmetric and higher spin generalizations. It would be nice to find a connection between these two realisations. We hope to report on this in future.

\vspace{1cm}
{\bf Acknowledgements}\\

We would like to thank Arindam Bhattacharjee, Dileep Jatkar, Wout Mebris and Turmoli Neogi for useful discussions. NB acknowledges hospitality at ICTP during the final stage of this work. Our work is partially supported by a SERB ECR grant, GOVT of India. Finally, we thank the people of India for their generous support for the basic sciences.

\appendix

\section{Conventions and Identities}\label{AppA}

In this appendix, we shall present our conventions. The tangent space metric $\eta_{ab}, a=0,1,2$ is flat and off-diagonal, given as
\begin{equation}
\nonumber
\eta_{ab}=\left(\begin{matrix}
0&1&0\\
1&0&0\\
0&0&1\\
\end{matrix}\right) .
\end{equation}
The space time coordinates are $u, \phi,r$ with positive orientation in the bulk being $du d\phi dr$. Accordingly the Levi-Civita symbol is chosen such that 
$\epsilon_{012}=1$.\\
The three dimensional Dirac matrices satisfy usual commutation relation  
$\{\Gamma_{a},\Gamma_{b}\}=2\eta_{ab}$ . They also satisfy following useful identities:
$$\Gamma_{a}\Gamma_{b}=\epsilon_{abc}\Gamma^{c}+\eta_{ab}\mathbb{I}, \hspace{21pt} (\Gamma^{a})^{\alpha}_{\beta}(\Gamma_{a})^{\gamma}_{\delta}=2\delta^{\alpha}_{\delta}\delta^{\gamma}_{\beta}-\delta^{\alpha}_{\beta}\delta^{\gamma}_{\delta}.$$
The explicit form of the Dirac matrices are chosen as,
\begin{equation}
\Gamma_0 = \sqrt{2}\left(\begin{array}{cc} 0 & 1 \\ 0 & 0\end{array} \right)\,, \qquad 
\Gamma_1 =\sqrt{2} \left(\begin{array}{cc} 0 & 0 \\ 1 & 0\end{array} \right)\,, \qquad
\Gamma_2 = \left(\begin{array}{cc} 1 & 0 \\ 0 & -1\end{array} \right)\,.
\end{equation}

All spinors in this work are Majorana and our convention for the Majorana conjugate of the fermions is given as, 
$$ \overline{\psi}_{\alpha i}=\psi^{\beta}_iC_{\beta\alpha}, \qquad
C_{\alpha\beta}=\epsilon_{\alpha\beta}=C^{\alpha\beta}=\left(\begin{matrix}
0&1\\
-1&0\\
\end{matrix}\right).\\ $$
Here $i=1,2$ is the internal index and $C_{\alpha\beta}$ is the charge conjugation matrix that satisfies 
$$C^{T}=-C, \hspace{11pt} C\Gamma_{a}C^{-1}=-(\Gamma_{a})^{T},\hspace{11pt} C_{\alpha \beta} C_{\beta \gamma} = - \delta_{\alpha \gamma} $$
For any traceless $2 \times 2$ matrix $A$, it can be shown that $C_{\alpha \beta} A^{\beta}_{\gamma} = (C\Gamma^a)_{\alpha\gamma} Tr[\Gamma_a A]$.\\

Other useful identities are:

\begin{equation}
\zeta\bar{\eta} = - \frac12 \bar{\eta}\, \zeta \, \mathbbm{1} - \frac12 (\bar{\eta}\Gamma^a \zeta)\Gamma_a, \qquad \bar\psi \G_a\,\eta=\bar\eta\,\G_a\,\psi\;,\qquad \bar\psi \G_a\,\epsilon= -\bar\epsilon\,\G_a\,\psi\;,
\end{equation}
where $\zeta,\psi,\eta$ are Grassmannian one-forms, while 
$\epsilon$ is a Grassmann parameter.

The generators of $sL(2,R)$ are considered as $\frac{\Gamma_i}{2}$. Further in Chevalley-Serre basis, they are given as,
\begin{equation}
E_{+} = \left(\begin{array}{cc} 0 & 1 \\ 0 & 0\end{array} \right)\,, \qquad 
E_{-} = \left(\begin{array}{cc} 0 & 0 \\ 1 & 0\end{array} \right)\,, \qquad
H = \left(\begin{array}{cc} 1 & 0 \\ 0 & -1\end{array} \right)\,.
\end{equation}
Thus, $E_{\pm}$ corresponds to positive(negative)root of the Cartan subalgebra and they are nilpotent.

\section{Terms in Reduced Action}\label{AppB}
For writing the phase space reduced action, we need to reduce various terms of \eqref{sa} in terms of Gauss Decomposed fields and further use the first class constraints relations of \eqref{1},\eqref{2},\eqref{3}. Below we note down the simplified forms of various terms:\\

{\bf Bosonic Terms :}\\
\begin{equation}
\operatorname{Tr}[2\mu\lambda^{-1}\lambda^\prime\lambda^{-1}\dot{\lambda}]=\mu\varphi^\prime\dot{\varphi}+2\mu(\dot{\sigma}\tau^\prime e^{-\varphi})+2\mu(\sigma^\prime\dot{\tau}e^{-\varphi})
\end{equation}
\begin{equation}
\operatorname{Tr}[-4\dot{\lambda}\lambda^{-1}F^\prime]=-4\Big[-\frac{\dot{\varphi}\theta^\prime}{2}-\sigma\dot{
	\tau}e^{-\varphi}\theta^\prime+\dot{\tau}\zeta^\prime e^{-\varphi}+\frac{\dot{\sigma}\eta^\prime}{2}-\frac{\sigma\dot{\varphi}\eta^\prime}{2}-\frac{\sigma^2\dot{\tau}e^{-\varphi}\eta^\prime}{2}\Big]    
\end{equation}
\begin{equation}
\operatorname{Tr}[-2(\lambda^{-1}\lambda^\prime)^2]=-2\Big(2\sigma^\prime\tau^\prime e^{-\varphi}+\frac{{\varphi^\prime}^2}{2}\Big)    
\end{equation}
\begin{equation}
\frac{2 \mu}{3} \int \operatorname{Tr}\left[\left(d \Lambda \Lambda^{-1}\right)^{3}\right]=\int du d\phi (-2\mu(\dot{\sigma}\tau^\prime e^{-\varphi})+2\mu(\sigma^\prime\dot{\tau}e^{-\varphi}))
\end{equation}
Rest of the terms in the action are scalar $2i\dot{a}C^\prime+\Bar{\mu}a^\prime\dot{a}-4(a^\prime)^2$ and they will remain as it is. Next we look at the fermionic terms.
\begin{equation}
\operatorname{Tr}[-2\left(\overline{d}_{1}^{\prime} \dot{d}_{2}+\overline{d}_{2}^{\prime} \dot{d}_{1}\right)]=2{d_1^-}^\prime \dot{d}_2^+-2{d_1^+}^\prime \dot{d}_2^-+2{d_2^-}^\prime \dot{d}_1^+-2{d_2^+}^\prime \dot{d}_1^-    
\end{equation}
\begin{equation}
\operatorname{Tr}[-2 i a^{\prime}\left(\overline{d}_{1 \alpha} \dot{d}_{2}-\overline{d}_{2 \alpha} \dot{d}_{1}\right.\Big)]=2ia^\prime d_1^-\dot{d}_2^+-2ia^\prime d_1^+\dot{d}_2^--2ia^\prime d_2^-\dot{d}_1^++2ia^\prime d_2^+\dot{d}_1^-    
\end{equation}
\begin{equation}
\begin{aligned}
\operatorname{Tr}[-2ia^\prime\dot{\lambda} \lambda^{-1}\left(d_{2} \overline{d}_{1}-d_{1} \overline{d}_{2}\right)]
&=(-2ia^\prime)[-2\sqrt{2}\dot{\tau}e^{-\varphi}d_2^-d_1^-+\sqrt{2}\dot{\sigma}d_2^+d_1^+-\sqrt{2}\sigma\dot{\varphi}d_2^+d_1^+\\&~~~~~~~~~~~~~~~~~~~~-\sqrt{2}\sigma^2\dot{\tau}e^{-\varphi}d_2^+d_1^+-\dot{\varphi}d_1^-d_2^+-\dot{\varphi}d_1^+d_2^-\\&~~~~~~~~~~~~~~~~~~~~-2\sigma\dot{\tau}e^{-\varphi}d_1^-d_2^+-2\sigma\dot{\tau}e^{-\varphi}d_1^+d_2^-]
\end{aligned}
\end{equation}
\begin{equation}
\begin{aligned}
\operatorname{Tr}[-4 \dot{\lambda} \lambda^{-1}\frac{1}{2}d_{1} \overline{d}_{2}^{\prime}]=& -2\Big[\frac{\dot{\varphi}}{2}d_1^+{d_2^-}^\prime+\sigma\dot{\tau}e^{-\varphi}d_1^+{d_2^-}^\prime-\sqrt{2}\dot{\tau}e^{-\varphi}d_1^-{d_2^-}^\prime+\frac{\dot{\sigma}}{\sqrt{2}}d_1^+{d_2^+}^\prime\\&-\frac{\sigma}{\sqrt{2}}\dot{\varphi}d_1^+{d_2^+}^\prime-\frac{\sigma^2}{\sqrt{2}}\dot{\tau}e^{-\varphi}d_1^+{d_2^+}^\prime+\frac{\dot{\varphi}}{2}d_1^-{d_2^+}^\prime+\sigma\dot{\tau}e^{-\varphi}d_1^-{d_2^+}^\prime\Big]
\end{aligned}
\end{equation}
\begin{equation}
\begin{aligned}
\operatorname{Tr}[-4 \dot{\lambda} \lambda^{-1}\frac{1}{2}d_{2} \overline{d}_{1}^{\prime}]=& -2\Big[\frac{\dot{\varphi}}{2}d_2^+{d_1^-}^\prime+\sigma\dot{\tau}e^{-\varphi}d_2^+{d_1^-}^\prime-\sqrt{2}\dot{\tau}e^{-\varphi}d_2^-{d_1^-}^\prime+\frac{\dot{\sigma}}{\sqrt{2}}d_2^+{d_1^+}^\prime\\&-\frac{\sigma}{\sqrt{2}}\dot{\varphi}d_2^+{d_1^+}^\prime-\frac{\sigma^2}{\sqrt{2}}\dot{\tau}e^{-\varphi}d_2^+{d_1^+}^\prime+\frac{\dot{\varphi}}{2}d_2^-{d_1^+}^\prime+\sigma\dot{\tau}e^{-\varphi}d_2^-{d_1^+}^\prime\Big]
\end{aligned}
\end{equation}
The trace terms are trivially zero.
We can further write all these terms in terms of newly defined fermions $d_1^N,d_2^N$. Finally combining them  and using the constraints, the reduced action looks like,
\begin{equation}
\begin{aligned}
I&=\frac{k}{4\pi}\int du d\phi\Bigg[\mu\varphi^\prime\dot{\varphi}+2\dot{\varphi}\theta^\prime+4\sigma\dot{\tau}e^{-\varphi}\theta^\prime-4\dot{\tau}\zeta^\prime e^{-\varphi}-2\dot{\sigma}\eta^\prime+2\sigma\dot{\varphi}\eta^\prime+2\sigma^2\dot{\tau}e^{-\varphi}\eta^\prime\\
&-{\varphi^\prime}^2+2i\dot{a}C^\prime+\Bar{\mu}a^\prime\dot{a}-4(a^\prime)^2+\dot{\varphi}d_2^{N+}(d_1^{N-})^\prime-\sqrt{2}\dot{\sigma}d_2^{N+}(d_1^{N+})^\prime-\dot{\varphi}d_2^{N-}(d_1^{N+})^\prime\\
&+\dot{\varphi}d_1^{N+}(d_2^{N-})^\prime-\sqrt{2}\dot{\sigma}d_1^{N+}(d_2^{N+})^\prime-\dot{\varphi}d_1^{N-}(d_2^{N+})^\prime+\sqrt{2}\sigma i\dot{a}(d_1^{N+})^\prime d_2^{N+}\\
&+\sqrt{2}\sigma(d_1^{N+})^\prime\dot{(d_2^{N+})}-2i\dot{a}(d_1^{N+})^\prime d_2^{N-}-2(d_1^{N+})^\prime\dot{(d_2^{N-})}-\sqrt{2}\sigma i\dot{a}(d_2^{N+})^\prime d_1^{N+}\\
&+\sqrt{2}\sigma(d_2^{N+})^\prime\dot{(d_1^{N+})}+2i\dot{a}(d_2^{N+})^\prime d_1^{N-}-2(d_2^{N+})^\prime\dot{(d_1^{N-})}\Bigg],
\end{aligned}
\end{equation}
using reduction relations this can be further simplified as,
\begin{equation}\label{ACTION}
\begin{aligned}
I&=\frac{k}{4\pi}\int dud\phi\Bigg[\mu\varphi^\prime\dot{\varphi}+\xi^\prime\dot{\varphi}-{\varphi^\prime}^2+2i\dot{a}C^\prime+\Bar{\mu}a^\prime\dot{a}-4(a^\prime)^2-\sqrt{2}\dot{\sigma}d_2^{N+}(d_1^{N+})^\prime\\&-\sqrt{2}\dot{\sigma}d_1^{N+}(d_2^{N+})^\prime+\sqrt{2}\sigma i\dot{a}(d_1^{N+})^\prime d_2^{N+}\\&+\sqrt{2}\sigma(d_1^{N+})^\prime\dot{(d_2^{N+})}-2i\dot{a}(d_1^{N+})^\prime d_2^{N-}-2(d_1^{N+})^\prime\dot{(d_2^{N-})}-\sqrt{2}\sigma i\dot{a}(d_2^{N+})^\prime d_1^{N+}\\&+\sqrt{2}\sigma(d_2^{N+})^\prime\dot{(d_1^{N+})}+2i\dot{a}(d_2^{N+})^\prime d_1^{N-}-2(d_2^{N+})^\prime\dot{(d_1^{N-})}\Bigg]
\end{aligned},
\end{equation}
where $\xi=2(\theta+\sigma\eta)+\big(d_1^{N+}d_2^{N-}+d_2^{N+}d_1^{N-}\big)$. Now notice one relation $d_1^{Ni}d_2^{Nj}=d_1^{i}d_2^{j},$ i.e. the product of redefined fermions are same as old ones. next to reduce the above action further we look at terms with $\dot a$. 
\begin{equation}
\begin{aligned}
I_{\dot a}&=\Bigg(\frac{k}{4\pi}\Bigg)\int dud\phi\Bigg[\sqrt{2}\sigma i\dot{a}(d_1^{N+})^\prime d_2^{N+}-\sqrt{2}\sigma i\dot{a}(d_2^{N+})^\prime d_1^{N+}-2i\dot{a}(d_1^{N+})^\prime d_2^{N-}+2i\dot{a}(d_2^{N+})^\prime d_1^{N-}\Bigg].
\end{aligned}
\end{equation}
Upto total derivatives in $\phi$ and using the reduction conditions we get,
\begin{equation}\label{adot}
\begin{aligned}
I_{\dot a}&=-2 i \Bigg(\frac{k}{4\pi}\Bigg)\int dud\phi \Bigg[- 2 \dot a e^{\phi}d_2^{N+} d_1^{N+}
+  (\dot a)^\prime \bigg((d_2^{N+}d_1^{N-}-d_1^{N+}d_2^{N-}) - \sqrt{2}\sigma d_2^{N+} d_1^{N+}\bigg)
\Bigg].
\end{aligned}
\end{equation}
Notice further that from all the above terms $N$ can be omitted. Further, the second term can be absorbed in redefinition of $c$ of \eqref{sa}. Next, we look at two fermion terms without $\dot a$ in \eqref{ACTION}. There are six such terms given as,
\begin{equation}\label{FA}
\begin{aligned}
I_{FF}&=\Bigg(\frac{k}{4\pi}\Bigg)\int dud\phi\Bigg[-\sqrt{2}\dot{\sigma}d_2^{N+}(d_1^{N+})^\prime-\sqrt{2}\dot{\sigma}d_1^{N+}(d_2^{N+})^\prime \\&+\sqrt{2}\sigma(d_1^{N+})^\prime\dot{(d_2^{N+})}+\sqrt{2}\sigma(d_2^{N+})^\prime\dot{(d_1^{N+})}-2(d_1^{N+})^\prime\dot{(d_2^{N-})}-2(d_2^{N+})^\prime\dot{(d_1^{N-})}\Bigg].
\end{aligned}
\end{equation}
The last two terms of the above expression are same as 3rd and 4th terms up to total $\phi, u$ derivatives. Thus we get,
\begin{equation}\label{FA1}
\begin{aligned}
I_{FF}&=\Bigg(\frac{k}{4\pi}\Bigg)\int dud\phi\Bigg[-\sqrt{2}\dot{\sigma}d_2^{N+}(d_1^{N+})^\prime-\sqrt{2}\dot{\sigma}d_1^{N+}(d_2^{N+})^\prime \\&+2\sqrt{2}\sigma(d_1^{N+})^\prime\dot{(d_2^{N+})}+2\sqrt{2}\sigma(d_2^{N+})^\prime\dot{(d_1^{N+})}\Bigg].
\end{aligned}
\end{equation}
The above four terms can be further simplified upto total derivatives and reduction conditions as,
\begin{equation}\label{FA2}
\begin{aligned}
I_{FF}&=\Bigg(\frac{k}{4\pi}\Bigg)\int dud\phi\Bigg[2 e^{\phi}\bigg( \dot d_1^{N+} d_2^{N+}+\dot d_2^{N+} d_1^{N+}\bigg)\Bigg].
\end{aligned}
\end{equation}
Let us now put the above equation \eqref{FA2} with the first term of \eqref{adot}, we get,
\begin{equation}\label{FA3}
\begin{aligned}
I&=2 \Bigg(\frac{k}{4\pi}\Bigg)\int dud\phi\Bigg[ e^{\phi}\bigg( \dot d_1^{N+} d_2^{N+}+\dot d_2^{N+} d_1^{N+}\bigg)+2 i \dot a e^{\phi}d_2^{N+} d_1^{N+}\Bigg]\\
&=2 \Bigg(\frac{k}{4\pi}\Bigg)\int dud\phi \ e^{\phi}\bigg( \dot d_1^{+} d_2^{+}+\dot d_2^{+} d_1^{+}\bigg).
\end{aligned}
\end{equation}
Finally we redefine $\chi_i= e^{\phi/2} \ d_i^{+}, \quad i=1,2$ and get,
\begin{equation}\label{FA4}
\begin{aligned}
I&=2 \Bigg(\frac{k}{4\pi}\Bigg)\int dud\phi \bigg( \dot \chi_1 \chi_2+\dot \chi_2 \chi_1\bigg).
\end{aligned}
\end{equation}
Combining all the terms we get the reduced action as in \eqref{LA}.
\section{Flat limit of Liouville Theory and its equivalent descriptions}\label{AppC}
In this appendix, we shall present some equivalent descriptions of Liouville theory in the "flat" limit. We shall mostly follow \cite{Barnich:2012rz, Donnay:2016zka}. A classical Liouville theory describes dynamics of a two dimensional scalar field $\phi$, such that when a two dimensional metric is scaled by $e^{2 \phi}$, the transformed metric has constant curvature $R$. The quantum Liouville action is given as 
\begin{equation}
S_L= \int d^2 x \sqrt{|g|}\bigg (- \frac{1}{2} g^{ab}\partial_a \phi\partial_b \phi + R \phi \frac{\gamma ^2+4}{2 \gamma} + \frac{\tilde \mu}{2 \gamma^2}e^{\gamma \phi}\bigg).
\end{equation}  
This is an interacting theory with $\gamma, \tilde \mu$ being constants. The above action in Hamiltonian form (that contains only one time derivative of field) on the Minkowskian cylinder (hence $R=0$)
with time coordinate time $u$, compact angular coordinate $\theta$\footnote{wrt lightcone coordinates of flat Minkowski space, $x_{\pm}= \frac{u}{l}\pm \theta$} and metric $ \eta_{\mu\nu}=$ diagonal $(-1,l^2)$, can be expressed as,
\begin{equation}
S_L= \int du d\theta \bigg( \pi \dot \phi - \frac{\pi^2}{2 }- \frac{\phi'^2}{2 l^2}-\frac{\tilde \mu}{2 \gamma^2}e^{\gamma \phi}\bigg),
\end{equation}  
here $\pi$ is conjugate momenta, $\dot \phi$ represents $u$ derivative and $ \phi'$ represents $\theta$ derivative. The action is invariant under two dimensional conformal transformations. We are interested in a large $l$ limit of this theory such that $$\phi= l \Phi, \pi = \frac{\Pi}{l}, \beta= \gamma l, \nu= \tilde \mu l^2$$ are fixed. The action in this limit looks like 
\begin{equation}
S_{FL}= \int du d\theta \bigg( \Pi \dot \Phi - \frac{\Phi'^2}{2}-\frac{\nu}{2 \beta^2}e^{\beta \phi}\bigg),
\end{equation} 
This is the flat limit of Liouville that preserves BMS$_3$ symmetry with zero $\mathcal{J}-\mathcal{J}$ central extension. One important point to note that this is a first order action and it does not have a second order counterpart. There are two equivalent description of the same theory that we shall list below. First one is a free field realisation given by,
\begin{equation}
S_{FL}= \int du d\theta \bigg( \pi_{\psi} \dot \psi - \frac{\pi_{\psi}^2}{2} + \frac{d}{d u}(\Phi \psi'- \frac{\sqrt \nu}{\beta}e^{\beta \Phi/2} \psi)\bigg)
\end{equation} 
where the fields are related by B$\ddot a$cklund transformations : $$\Pi=\psi'- \frac{\sqrt \nu}{2}e^{\beta \Phi/2} \psi, \quad \pi_{\psi}= \Phi'+ \frac{\sqrt \nu}{\beta}e^{\beta \Phi/2}. $$
The second realisation is given as,
\begin{equation}
S_{FL}= \frac{k}{4 \pi}\int du d\theta (\xi^\prime\dot{\varphi}-{\varphi^\prime}^2)
\end{equation} 
where field transformations are $$\beta \Pi=  \xi ' - (\log \sigma)' \xi, \quad \beta \Phi= 2 \varphi- 2 \log \sigma - \log \frac{8}{\nu},$$ 
with $\beta^2= 32 \pi G, \sigma'= \sqrt 2 e^{\varphi}.$ The second description arises as the reduced phase space description of $SL(2,R)$  chiral WZW model with appropriately constrained (due to specific asymptotic boundary conditions of fields at null infinity) global currents.

\bibliography{bms3}
\bibliographystyle{jhep}

\end{document}